\documentclass[aps,prd,onecolumn,12pt,preprintnumbers]{revtex4-1}

\usepackage{hyperref}

\usepackage{float}

\usepackage{tikz}

\usepackage{graphicx}
\usepackage{epstopdf}

\begin{document}

\preprint{HDP: 18 -- 01}

\title{\centerline{Whither Tone Ring Ring?}}

\author{David Politzer}

\email[]{politzer@caltech.edu}

\homepage[]{http://www.its.caltech.edu/~politzer}

\altaffiliation{\footnotesize 452-48 Caltech, Pasadena CA 91125}
\affiliation{California Institute of Technology}

\date{May 21, 2018}

\bigskip

\bigskip

\begin{figure}[h!]
\includegraphics[width=4.5in]{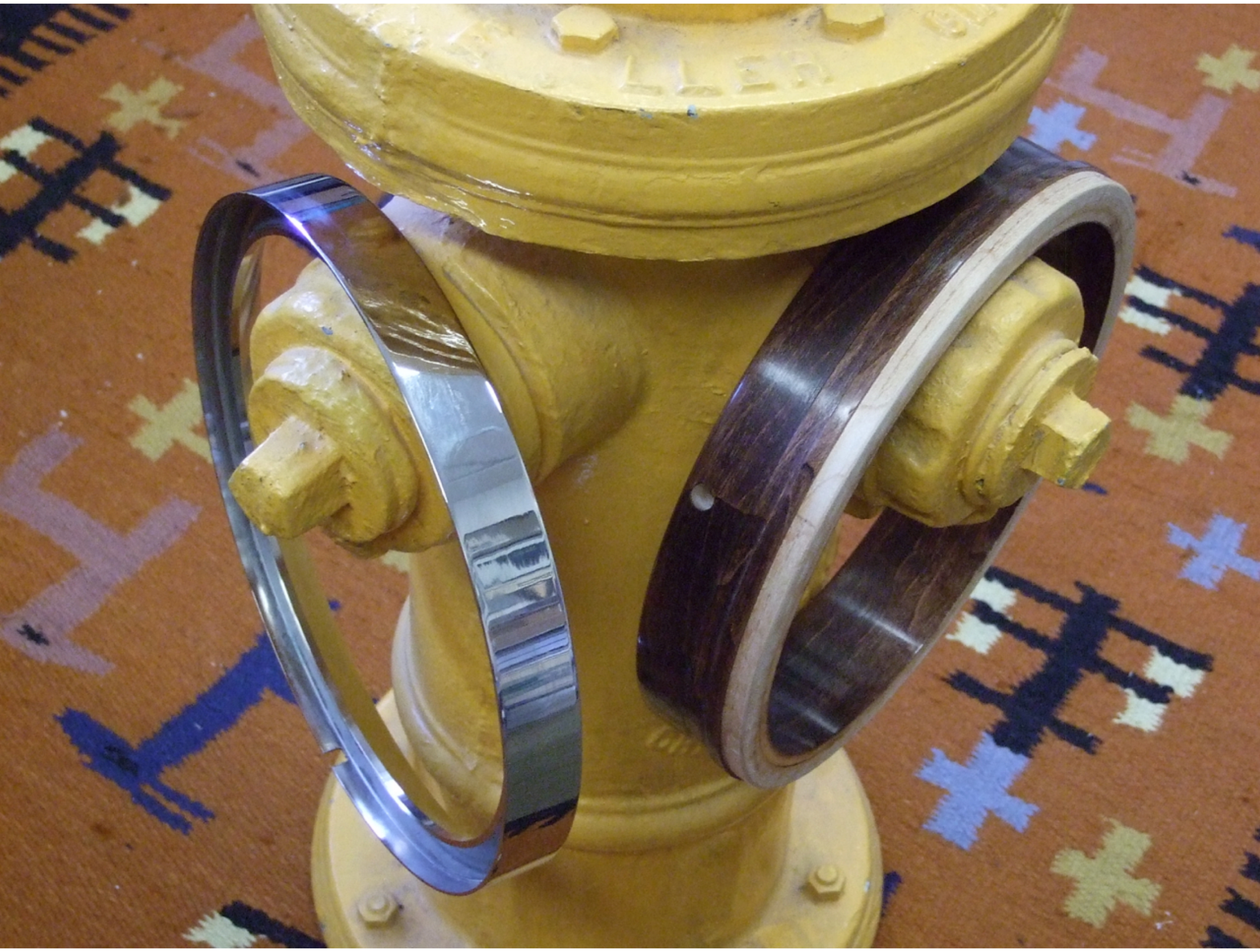}
\end{figure}

\bigskip

\begin{abstract}

An extremely simple model captures the essence of the interaction of a banjo tone ring with the wood rim.  The large scale, low frequency resonances of the assembled system are related to the weights and resonant frequencies of the tone ring and rim separately.  Very crude measurements satisfy the derived relations within about 5\% for the lowest frequency modes and give qualitative agreement for the next ones on a particular, heavy-tone-ring resonator banjo.  The two combined sub-systems become increasingly independent for higher frequency, shorter-lived modes.   Nevertheless, the ringing sounds of the struck individual parts, which dominate the perception of their pitch and sustain, are related by the simple model to the sound when the parts are struck when combined into one. 

\end{abstract}

\maketitle{{\centerline{\large \bf Whither Tone Ring Ring?}}

\section{Introduction}

``Ring, Ring the Banjo," wrote Stephen Foster back when banjos had few if any metal parts.  But over the decades, metal was often added.  Banjos can have metal tension rings, bracket bands, flanges, and, very notably, tone rings. Models were once named ``Silver Chime" and ``Silver Bell."  Pick up a tone ring and whack it.  It rings out, long and sweet, like a chime or bell.  Whack an unadorned wood rim, and it sounds like a wood-block.  Where do those chime tones go when the instrument is assembled? Tone rings provide the most extreme examples.  The frequencies of their clear, chiming solo sounds simply do not show up as enhancements or suppressions in the sound of the played instrument.  Clearly, the long ring sound of the tone ring by itself is damped out by its contact with the other parts.  But what happens to the resonant frequencies of that assiduously wrought metal ring?

The photo on the title page shows the nickel-plated tone ring and wood rim of a 2004 Deering Sierra. Their designed fit is very snug, even without the down-pressure of the head.  I spliced together a sound file with one tap for each ring: the tone ring, the wood rim, and the two combined:

\medskip

\centerline{\href{http://www.its.caltech.edu/~politzer/ring-taps/one-tap-each.mp3}{http://www.its.caltech.edu/\url{~}politzer/ring-taps/one-tap-each.mp3} }
\medskip
\noindent (Copy the link or just click.  They're hung by a thread; so it's about 30 seconds long; listen to the end.)

Certainly, a tone ring's mechanical properties contribute to the sound.  I present here a very simple picture of how those properties effect the sound of the combined tone-ring-and--wood-rim assembly.  The relevant parameters can be estimated from the weights of the ring and rim and from the resonant frequencies of their recorded, individual sounds and their decay times.  In practice, this analysis only applies to the low frequency, large scale motions.  Higher frequency motions deviate from the underlying simple assumptions.  And this model has nothing to do with what is likely a tone ring's most important job: to improve the reflection of high frequency head vibrations back onto the head from its edge.  Discerning players are inordinately fussy about the particulars, and most fine details of design and timbre are beyond the present discussion.  But, even if your tone ring has ball bearings, scallops, flanges, or other intricate designs, its largest scale motions and most prominent tap sounds will behave as described here.

\section{The Model}

The crucial observation is that, once it's installed, a tone ring is constrained to move with the rim it sits on. In some cases, such as the 2004 Deering Sierra used in the measurements described below, there's a very snug fit.  That tone ring has a flange that fits tightly around a carefully turned rim.  The fit is so snug that the tone ring and rim sound out as one when assembled and tapped.  However, even a ring that simply sits on top of the rim will have down-pressure from a properly tightened head, ensuring that there's no appreciable relative motion of the ring with respect to the rim --- at least along their surfaces of contact.  In fact, a head, tension ring, and hooks tightened to playing tension provide a much better approximation to the idealized notion that the tone ring and rim have no relevant, relative motion.  However, a fully assembled pot would be far more difficult to analyze.

That the two parts are constrained to move together is the central observation and also the origin of the important caveats.  Even if there were absolutely no slipping at the two parts' contact, there will certainly be some amount of independent motion of material somewhat distant from those contacts.  Nevertheless, the lowest frequency ``normal modes" (resonant motions) --- and a whole series of their higher frequency relatives --- do have this lock-step, co-moving aspect to the assembled parts.  These are the ``ring" modes.

\subsection{Ring Modes}

A thin, solid, circular ring has vibrational modes with alternating-sign displacements from equilibrium that are evenly spaced around the circumference.  The number of nodes is even and greater than or equal to four.  For a given mode, the displacements can be radial, i.e., in the plane of the ring; ``vertical," i.e., out of the plane (which is the plane of the head for a banjo); or torsional, twisting around the ring's centerline.\cite{ring-modes}  The restoring force is Young's modulus, i.e., the opposition of the material to stretching and compression.  There are no simplifying assumptions or limits in which the ring is a soluble problem.  There is no analytic solution --- no formula, simple or otherwise.  It was not that long ago that PhD theses in mechanical engineering focused on improving approximation techniques for computing ring modes and testing them against vibrations of rings and cylinders.

But, for small amplitudes, each mode behaves like a harmonic oscillator --- even if we do not know its exact shape along the ring circumference or the relation of the frequency to the mechanical properties of the material or the geometry of the ring cross section.

\subsection{The Oscillator Equation per Mode}

A particular mode, say with displacements in the radial direction in the ring plane, has a frequency $\omega$ (in radians) that we can measure, an effective mass $m$ or inertia, and an effective spring constant $k$ or restoring stiffness, all related by $\omega^2 = k/m$.

\begin{figure}[h!]
\includegraphics[width=2.5in]{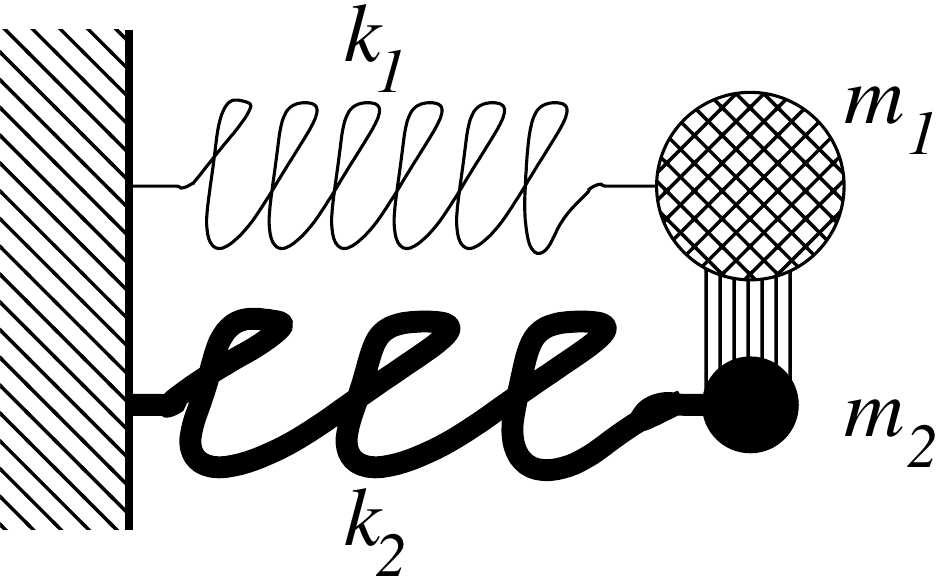}
\caption{oscillators constrained to move together}
\end{figure}
If we have two oscillators with frequencies $\omega_1$ and $\omega_2$, masses $m_1$ and $m_2$, and spring constants $k_1$ and $k_2$ and then constrain them to move together as in FIG.~1, then the mass of the combined, constrained system oscillator is $m_1+m_2$ and the spring constant $k_1+k_2$.  Then the frequency of the combined, constrained system can be expressed as

\smallskip
\centerline{\large $ \omega^2 =  {\omega_1^2 \over {1 + m_2/m_1}} + { \omega_2^2  \over{1 + m_1/m_2}} $ .}
\smallskip
\noindent In particular, we need not know the $k$'s and only need know the ratio of the $m$'s to get the combined frequency from the two separate frequencies.  (It might also help to note that this is a weighted average.  If $m_1=m_2$, then $\omega^2$ is the average of the two individual $\omega^2$'s.)

The frequencies can be measured by listening (with sound software including spectrum analysis).  Identifying which mode is which, e.g. distinguishing radial from vertical motion, is a practical question, discussed below with some actual measurements.  But what about the mass ratios?

\subsection{The Inertia--Density Assumption}

When considering two different rings of different materials and cross section geometry, I will assume that

\centerline{\large ${m_1 \over m_2} = {M_1 \over M_2}$}
\noindent where $m_{1,2}$ are the inertias of the corresponding modes and $M_{1,2}$ are the total masses of the corresponding rings.  $M_1$ and $M_2$ are easily measured.  So the mode ratio is assumed to be the same for all modes.

If $m_{1,2}$ were simply the same common ratio of the mass between nodes, then the relation to $M_1/M_2$ would be exact.  Such is, in fact, the case with some simple, exactly soluble systems, such as ideal strings and membranes.  In those cases, the spatial shape of the modes for a given node number is universal.  (For the string, it's sinusoidal.)  However, the effective mass or inertia of a particular mode of a ring, the effective ``spring constant" for that mode, and the mode spatial shape all depend on the distribution of material about the neutral strain plane.  As a result, the spatial shapes for the normal modes for a given number of nodes of the ring, the rim, and the combined system may be slightly different.   A small amplitude, ``uniform, same-thin-shape ring" approximation might restore the equality.  In any case, that's what is in the following.

\section{Sound Measurements}

I hung a tone ring, wood rim, and the assembled combination from a thread; tapped with a piano hammer; and recorded the sound with a USB microphone.  The parts were from a 2004 Deering Sierra, which has a very snug fit.  The sound of taps at all positions and from all angles suggested that there was very little independent motion of the two rings when assembled, even without the pressure of a head.  Of course, some sound variation could be elicited by tapping at some particular locations on the tone ring surface.  Without the down pressure of tension hooks and head, there may well be a small amount of motion between the parts when assembled (contradicting the assumption of the simple model).  But adding the head and tension ring complicates the system considerably.

Tapping in the radial direction produced different pitches from taps in the vertical direction.  I realized that I could separate the modes even more clearly if I listened closely (within ${1 \over 2}''$) at the location of an expected anti-node (maximum) and in the right direction (radially or vertically from the surface).  So that's how I positioned the microphone in successive runs. 

The suspending thread and gravity broke the rotational symmetry inherent in the rings.  In particular, radial modes whose nodes are not at 12:00 and 6:00 o'clock involve raising and lowering the center of mass as the ring vibrates.  Working against gravity would break rotational symmetry and raise the frequency of some modes.  So I put nodes there (at least for the lowest mode) by tapping at 10:30 --- and listening at 1:30.

\begin{figure}[h!]
\includegraphics[width=6.5in]{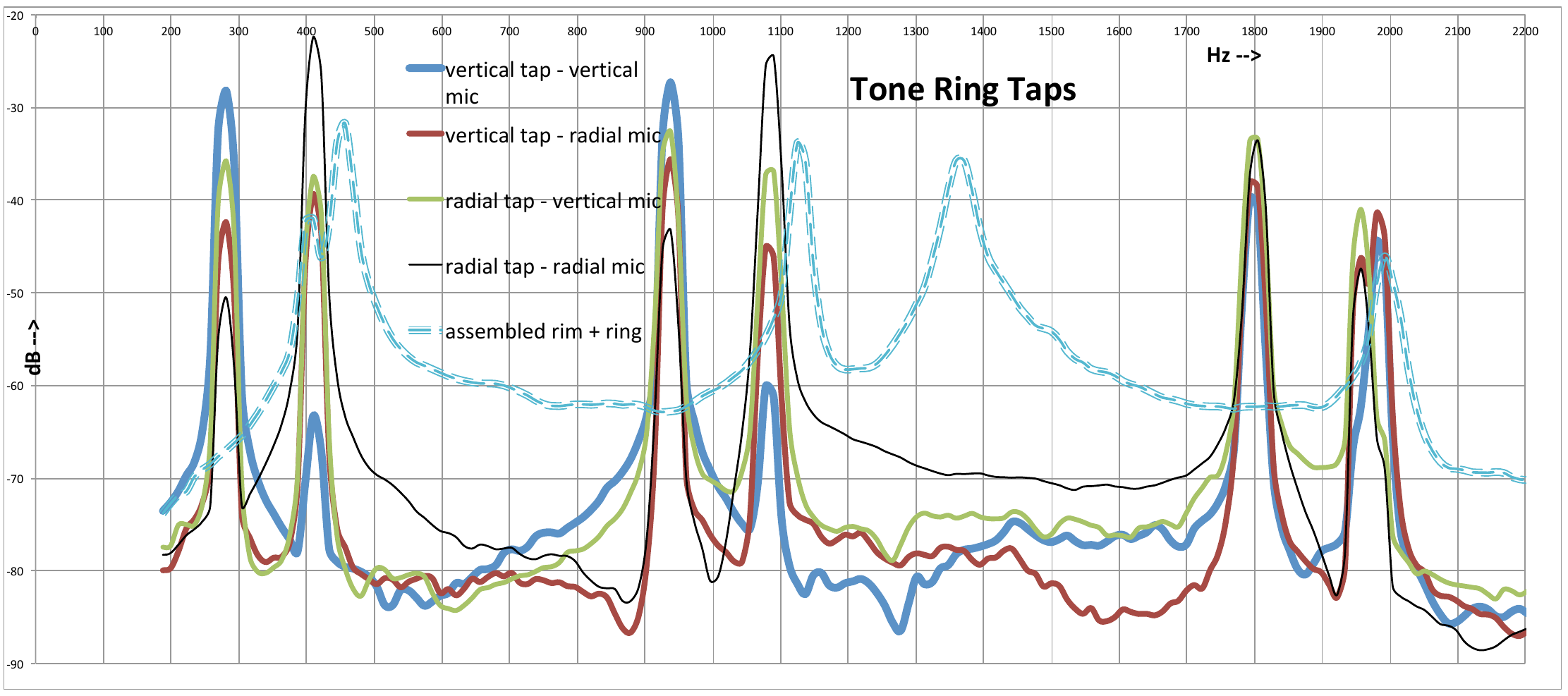}
\includegraphics[width=6.5in]{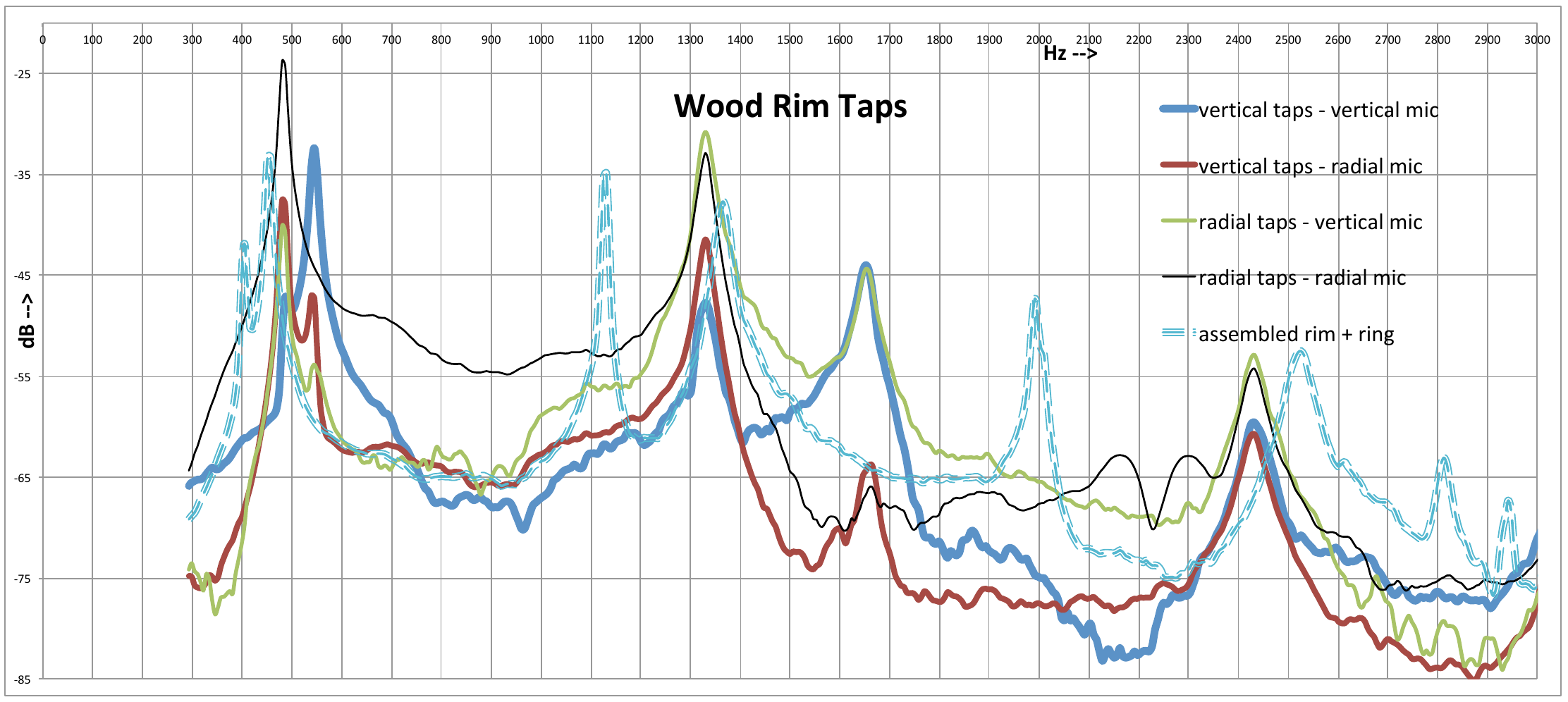}
\caption{Tone ring, wood rim, and combined system spectra.  Be aware that neither vertical nor horizontal scales are quite the same in the two graphs.  Most significantly, one goes to 2200 Hz, the other to 3000 Hz.}
\end{figure}

FIG.~2 shows the first few resonances of the separate rings and the combined system.  The various curves are labeled by the direction of the tap and the location and direction of the microphone.  This allowed an unambiguous  identification of radial versus vertical motion, at least in a few cases.  The lowest two resonances for the tone ring in FIG.~2 are the clearest example.  Vertical taps with a vertically oriented microphone produced a much stronger response at the lower of the two frequencies.    (Again, ``vertical" means perpendicular to the plane of the head.)  Radial taps with a radially oriented microphone produced exactly the same two frequencies, but the higher one was much stronger.  This implies that the lower frequency corresponds to the lowest (4 node) mode in which the tone ring vibrates out of its equilibrium plane.  The second peak is the 4 node resonance whose motions are radially in and out.

The $3^{\text{rd}}$ and $4^{\text{th}}$ peaks of the tone ring spectra show the same relation of intensity and direction, implying again that the $3^{\text{rd}}$ is a vertical motion and the $4^{\text{th}}$ a radial motion. $5^{\text{th}}$ and $6^{\text{th}}$ do not show any such distinction.  Perhaps they are neither, but, rather, the lowest torsional modes.  Alternatively, as the frequency increases, the number of nodes becomes larger for the radial and the vertical resonances, and the spacings around the circumference becomes smaller.  With a series of crudely positioned taps and a finite size, hand-held microphone, it would be increasingly difficult to pick up that sort of difference.  And perhaps, I was just too close to one of the nodes of the 8 node resonances.

The recorded tap spectrum of the assembled ring and rim is included in both graphs.  The four possible configurations of positioning showed only very slight systematic differences.  So, for these particular graphs, I combined them into a single line.  Note that it seems to have a mind of its own relative to the spectra of its parts.  When the constrained oscillator model is applied to the four cases where the motion has been unambiguously identified, the relation of the spectra make some sense.

\section{Plugging in the Numbers}

I weighed the two items at the Post Office: $M_{\text{ring}}$ = 3 lb 0.5 oz and $M_{\text{rim}}$ = 1 lb 1.6 oz.  That gives 
$M_{\text{ring}}/M_{\text{rim}} = 2.76$ .

I used Audacity to do the recordings and spectrum calculations that are plotted in FIG.~2.  From the actual data files, I estimated the following frequency values for the first three rows and computed a value for the combined system using the simple theory, the ring and rim weights, and the first two rows' frequency data.

\begin{center}
\begin{tabular} {c | c | c | c | c}
frequency in Hertz  & n=4 radial & n=4 vertical & n=6 radial & n=6 vertical \\
\hline  
ring & 412 & 280 & 1083 & 936 \\  
rim & 485 & 554 & 1333 & 1655 \\ 
combo - measured & 454 & 404 & 1140 & 1366 \\ 
combo - theory & 433 & 373 & 1155 & 1171 
\end{tabular}  
\end{center}

\bigskip

\section{Combined Decay Times}

The free decay of the linearly damped harmonic oscillator, $m \ddot x = -kx -b\dot x$, goes, in time, like e$^{-\Gamma t} \times$ a sinusoidal $t$ variation.  The exponential free decay rate satisfies $\Gamma = b/2m$. 
\begin{figure}[h!]
\includegraphics[width=2.5in]{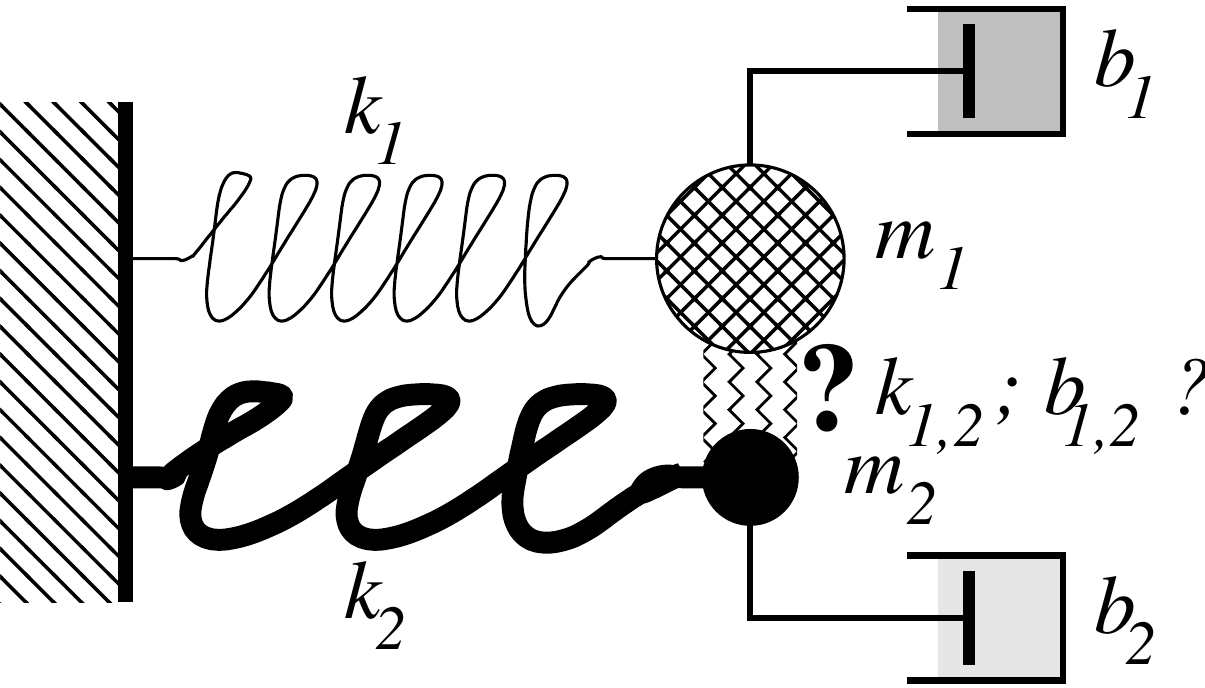}
\caption{approximately constrained oscillators, including damping  }
\end{figure}
\noindent  Damping in mechanical systems is rarely linear, but the $-b\dot x$ term often gives a good account of the gross behavior, especially if the damping is weak, i.e., if the system goes through many
 oscillations before the amplitude decreases appreciably.   If two oscillators are constrained to move together (as depicted in FIG.~3 --- ignoring any relative motion or $b_{1,2}$ for now), then 

\medskip

\centerline{\large $\Gamma = {\Gamma_1 \over {1+m_2/m_1}} + {\Gamma_2 \over {1+m_1/m_2}}$}

\bigskip

In principle, the width (in frequency) of a resonance peak is proportional to $\Gamma$ for the relevant oscillation.  That is apparent in FIG.~2.  The long-lived resonances of the tone ring are much sharper as functions of frequency than the short-lived ones of the wood rim and combined system.  In practice, various methods of measuring or calculating the widths can make them appear larger than they actually are.  And that is the case here, mostly because I used what was conveniently packaged with Audacity to do the spectrum calculation.  Looking at the recorded sound amplitude as a function of time is a more direct method to determine the decay rates.

FIG.~4 is a screen shot from Audacity, derived from a recording of a series of taps on the wood rim.  The horizontal scale is measured in seconds.  The vertical scale is in decibels.  In particular, the vertical scale is logarithmic in the recorded microphone voltage.  Most importantly, the original recording has been processed through a narrow band-pass filter (easy to specify in Audacity) centered on what was identified as the lowest vertical rim resonance, i.e., 554 Hz.
\begin{figure}[h!]
\includegraphics[width=5.0in]{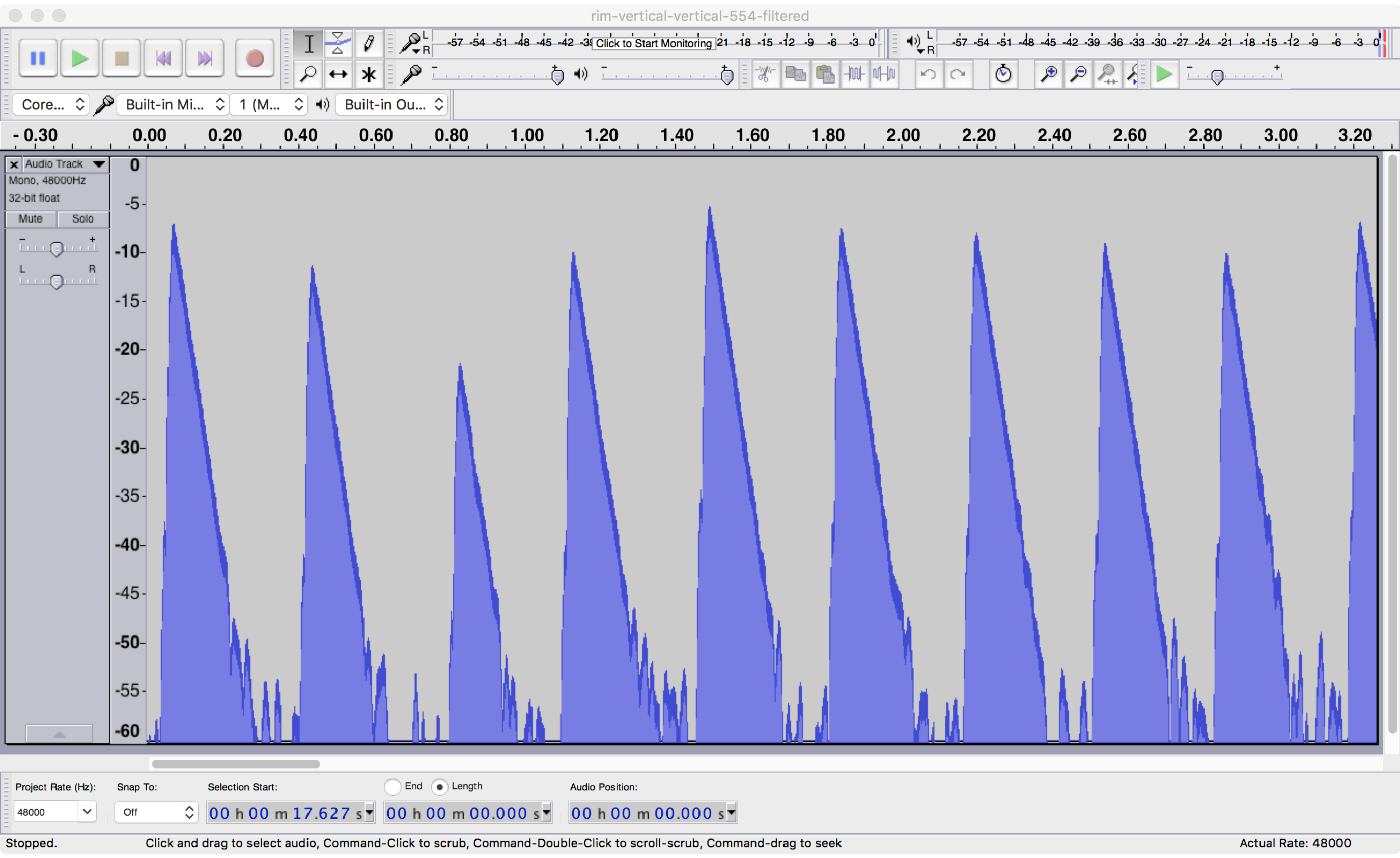}
\caption{screen shot from Audacity of successive rim taps, narrow-band-filtered around the lowest vertical resonance and plotted as dB {\it versus} time in seconds}
\end{figure}
\noindent Of the successive taps, some were louder and some softer.  But all show the same approximately exponential decay rate in time, i.e., the downward slope --- something like 35 dB in 0.15 seconds.  To the extent that all decays are roughly linear on the log scale and all have the same slope, the linear model for dissipation is a reasonable representation.  The Audacity plots can be turned into numbers by expanding the horizontal time scale.   Under closer scrutiny with an expanded time scale, it is clear that the decays are not exactly single exponentials.  The metal ring, for example, produces clear beats, evident even from a single, unfiltered recording.\cite{tap-mp3}  The combined ring-and-rim system also exhibits beats.  (Beats in the rim taps are just not as prominent.)  The beats result from two frequencies that are too close together to be resolved by the available spectrum calculation or band-pass filter.  However, they are totally expected.  If the ring and rim were perfectly rotationally symmetrical, every mode would be ``doubly degenerate," the physics term for two distinct motions having the same frequency.  Any slight deviation from perfectly round produces a splitting in their frequencies.  Both are produced by a single tap, and they produce a time-dependent interference in their combined sound; it throbs at the difference frequency. 

Perhaps more significantly, when examined very closely using expanded time scales, the overall decays are simply not straight lines in the log plots.  So what I report below are just reasonable approximations to the dominant behavior, i.e., the slope values that at first seem so evident in representations like FIG.~4.  (The numbers are good to about 10\%.)

\begin{center}
\begin{tabular} {c | c | c}
$\Gamma$ in dB/sec  &  n=4 radial & n=4 vertical \\
\hline  
ring & 16 & 1.1  \\  
rim & 219 & 259  \\ 
combo - measured & 154 & 93  \\ 
combo - theory & 70 & 68 
\end{tabular}  
\end{center}

\medskip

The combined system decays faster than predicted by the simple model of constraining the two parts to move together.  Were the measured decay slower than the model prediction, it would be a conundrum.  Faster simply suggests that there is some small relative motion in some or all of their surfaces of contact as the combined system oscillates.  Friction at the interface would increase the decay rate.  The simplest model of masses and springs that could represent the combined motions and include this relative motion would have two masses, each with their own spring and damping, {\it and} a spring and damping term for their relative motion.  This is the system of two coupled, damped oscillators.  An explicit, closed form solution exists and can be written down in terms of the various mechanical parameters.  However, no one ever writes it down because it's just too complicated.  Some properties can be highlighted by looking at particular, simplifying limiting values.\cite{coupled-damped}

The two-coupled-oscillator system has two distinct decaying modes.  In the lower frequency mode, the two move {\it more-or-less} together, while in the higher frequency mode their motions are {\it more-or-less} opposite.  The limiting values appropriate to the present situation would have the spring constant and damping coefficient of the relative motion be much larger than the individual ones.  That gives a wide separation of the two mode frequencies and gives a contribution to both modes' damping from the coupled damping.  The resulting picture is that the small relative motion of the ring and rim adds significantly to the decay rates of their combined resonances, while the frequencies are still described roughly by the original, constrained-motion model.  Going any further would require examining the relative motions and friction.  I will not do that here.

\section{The Lesson}

The calculated numbers would likely have worked out better had the tone ring and rim been more firmly attached, as is assured in an assembled pot.  However, any simple implementation would preclude ringing the ring/rim system by itself.

A tone ring certainly contributes to the mechanical properties of the pot, but it does not ring out on its own.  It contributes stiffness and mass that effect the pot motion.  This is especially true for its lowest frequencies and their sustain, the features that dominate our perception of its ring sound when it is by itself.  Not included in the presented model are smaller scale motions of the tone ring that effect its response to higher frequencies.  These, in turn, modulate what the head can and cannot do in turning string vibrations into sound.


\bigskip

\begin{thebibliography}{99}

\bibitem{ring-modes}These are modes that leave the circumferential length unchanged ``to lowest order." There is no return force or vibration with $n=2$ among transverse ring modes.  That's why there are at least four nodes.  Longitudinal waves, i.e., stretching and contracting along the circumferential direction, are conventional sound waves within the material and have much higher frequencies for a given wavelength.  For example, the two-node longitudinal resonance in an $11''$D brass ring should have a frequency of about 2700 Hz, based on the bulk speed of sound.

\bibitem{tap-mp3}as in the aforementioned \href{http://www.its.caltech.edu/~politzer/ring-taps/one-tap-each.mp3}{http://www.its.caltech.edu/\url{~}politzer/ring-taps/one-tap-each.mp3}.

\bibitem{coupled-damped}I reviewed what happens when the two oscillators, separately, are nearly degenerate, and the coupling and all dampings are weak in comparison.  A plucked string and a ring by itself are interesting examples where those approximations apply.  One finds that there may or may not be beats, and the decay may be a single exponential or the sum of two with different rates.  See \href{http://www.its.caltech.edu/~politzer}{D.~Politzer, {\it Zany strings and finicky banjo bridges}, http://www.its.caltech.edu/\url{~}politzer}, HDP: 15-- 01; scroll down to July 2014; or {\it The plucked string: an example of non-normal dynamics}, also at \href{http://www.its.caltech.edu/~politzer} {http://www.its.caltech.edu/\url{~}politzer}, July 2014 or American Journal of Physics {\bf 83} 403 (2015).


\end{thebibliography}
\end{document}